# Topology-Aware Focal Loss for 3D Image Segmentation

Andac Demir
Novartis
andac.demir@novartis.com

Elie Massaad
Mass. General Hospital
emassaad@mgh.harvard.edu

Bulent Kiziltan
Novartis
bulent.kiziltan@novartis.com

## Abstract

*The efficacy of segmentation algorithms is frequently compromised by topological errors like overlapping regions, disrupted connections, and voids. To tackle this problem, we introduce a novel loss function, namely Topology-Aware Focal Loss (TAFL), that incorporates the conventional Focal Loss with a topological constraint term based on the Wasserstein distance between the ground truth and predicted segmentation masks' persistence diagrams. By enforcing identical topology as the ground truth, the topological constraint can effectively resolve topological errors, while Focal Loss tackles class imbalance. We begin by constructing persistence diagrams from filtered cubical complexes of the ground truth and predicted segmentation masks. We subsequently utilize the Sinkhorn-Knopp algorithm to determine the optimal transport plan between the two persistence diagrams. The resultant transport plan minimizes the cost of transporting mass from one distribution to the other and provides a mapping between the points in the two persistence diagrams. We then compute the Wasserstein distance based on this travel plan to measure the topological dissimilarity between the ground truth and predicted masks. We evaluate our approach by training a 3D U-Net with the MICCAI Brain Tumor Segmentation (BraTS) challenge validation dataset, which requires accurate segmentation of 3D MRI scans that integrate various modalities for the precise identification and tracking of malignant brain tumors. Then, we demonstrate that the quality of segmentation performance is enhanced by regularizing the focal loss through the addition of a topological constraint as a penalty term.*

## 1. Introduction

In the domain of image segmentation, several deep learning approaches have been proposed. Notable among them are U-Net [38], Mask R-CNN [19] and SegNet [2]. However, these techniques have limited scalability when applied to large 3D volumes. In order to overcome this limitation, modifications such as multi-scale architectures [45] and attention mechanisms [33] have been suggested. Nevertheless, these approaches still suffer from deficiencies such as limited receptive field, generalization and scalability. Specifically, they can only capture a limited context around the target voxel, which hinders their effectiveness. Although larger kernel sizes or dilated convolutions can address this issue, it comes at the cost of computational complexity. Another challenge is their sensitivity to data variability, which can be addressed through data augmentation and transfer learning. Lastly, the limited scalability of these techniques to larger 3D volumes is attributed to the need for multiple downsampling and upsampling steps that result in information loss and reduced accuracy.

Recent advances in image segmentation using convolutional neural networks (CNNs) have shown remarkable results. However, the performance of these networks can be further enhanced by incorporating topological data analysis (TDA) tools [21, 22, 32]. TDA tools are useful for analyzing high dimensional data, and can capture important geometric and structural information about the data.

One way to incorporate TDA into CNNs is to add topological layers to the network architecture [28]. These layers can capture the persistence of the homology groups of the input data using a persistence diagram layer. Alternatively, topological features can be computed using TDA methods like persistent homology or mapper, and used as input to the CNN, thereby improving the robustness and accuracy of the network.

In addition, topological loss can be integrated into the training of a neural network to improve its performance for image classification tasks [22, 30]. This loss function measures the topological consistency of the network output with the input data, encouraging the network to learn features that are topologically consistent. This is achieved by modifying the standard cross-entropy loss function to include a term that measures the topological consistency of the network output with the input data using again TDA methods like persistent homology or mapper.

Machine learning methods have advanced to the point where brain tumors can be segmented, and tumors' genetic and molecular biology can be predicted based on MRI. [39]

However, none of the attempted approaches were clinically viable because of limited performance. [6]

Existing methods provide limited capabilities to harness the variability in tumor shape and texture that characterize different types of diffuse gliomas. Hence we propose using topological features computed by persistent homology to identify malignant entities. We hypothesize that topological features contain useful complementary information to image-appearance based features that can improve discriminatory performance of classifiers.

### 1.1. Our contributions

- We introduce a novel loss function, TAFL, that combines the Focal Loss with a topological constraint term based on the Wasserstein distance between the ground truth and predicted segmentation masks' persistence diagrams.

- We construct persistence diagrams from filtered cubical complexes of the ground truth and predicted segmentation masks and utilize the Sinkhorn-Knopp algorithm to determine the optimal transport plan between the two persistence diagrams.

- We evaluate TAFL by training a 3D U-Net with the MICCAI Brain Tumor Segmentation (BraTS) challenge validation dataset, which requires accurate segmentation of 3D MRI scans for the precise segmentation of malignant brain tumors.

- We demonstrate that the quality of segmentation performance is enhanced by regularizing the focal loss through the addition of a topological constraint as a penalty term. Overall, our approach provides an effective solution to the challenge of topological errors in segmentation algorithms.

## 2. Related Work

### 2.1. Deep Learning in Brain Tumor Detection

Central nervous system (CNS) tumors are among the most fatal cancers in humans of all ages. Gliomas are the most common primary malignancies of the CNS, with varying degrees of aggressiveness, morbidity, and mortality. Currently, most gliomas require tissue sampling for diagnosis, tumor grading, and identification of molecular features to targeted therapies. Accurate segmentation of gliomas is important for surgeons because it can help them better understand the tumor's location, shape, and size, which can inform treatment planning and improve surgical outcomes.

Upon clinical presentation, patients typically undergo imaging followed by biopsy alone or biopsy with resection for diagnosis and determination of histopathologic classification, tumor grade, and molecular markers. [46] Differentiation of low-grade gliomas (LGGs; grade II) from high-grade gliomas (HGGs; grades III, IV) is critical, as the prognosis and thus the therapeutic strategy could differ substantially depending on the grade. [11, 26] Tissue biopsy, while considered the diagnostic gold standard, is invasive and can sometimes be high-risk or inconclusive. The Cancer Genome Atlas (TCGA) report suggests that only 35% of biopsy samples contain sufficient tumor content for appropriate molecular characterization. [43] Hence, robust non-invasive approaches are desired for accurate diagnosis and monitoring of tumor evolution.

The state-of-the-art models in brain tumor segmentation are based on the encoder-decoder architectures, with U-Net-like architectures [40] (basic U-Net [14, 38], UNETR [18], Residual U-Net [25], and Attention U-Net [33]), were among top submissions to the BraTS challenge. Among factors influencing models performance like architecture modifications, training schedule, and integration of multiple MRI sequences, the size of brain tumor datasets remains a major limiting factor. Additionally, based on earlier successes for classification of tumor grades (high grade vs. low grade), a few studies explored the determination of glioma mutation status (O6-Methylguanine-DNA methyltransferase (MGMT) promoter methylation status) using radio-genomics features. Of importance, analysis of MGMT promoter methylation is essential to predict prognosis and treatment response. As a result, TDA has the potential to support future research in MRI-based deep learning to reveal clinically significant tumor features and aid clinical decision making.

According to several ablation studies run by [14] in order to select the most optimal CNN architecture for 3D medical image segmentation, UNet [38] attains better tumor segmentation performance as measured by Dice score compared to Attention UNet [33], Residual UNet [20], SegResNetVAE [31] and UNetR [18]. Hence, we adopt the U-Net architecture to encode each MRI scan image into a latent-space representation.

### 2.2. TDA in Computer Vision

Persistent homology, which is the primary tool in topological data analysis (TDA), has demonstrated remarkable efficacy in pattern recognition for image and shape analysis. Over the past two decades, numerous works in a variety of fields have utilized persistent homology for a range of applications, including the analysis of hepatic lesions in images [1], human and monkey fibrin images [7], tumor classification [12, 35, 36], fingerprint classification [16], diabetic retinopathy analysis [13, 15], the analysis of 3D shapes [42], neuronal morphology [24], brain artery trees [5], fMRI data [37, 44], and genomic data [9]. Of note, the TDA Applications Library [17] catalogs hundreds of compelling applications of TDA in various fields. For a comprehensive survey of TDA methods in biomedicine, see the survey [41].

On the other hand, TDA has also been successfully applied to the problem of tumor classification in several works, such as [12, 35, 36]. In [36], the authors employed TDA techniques to study histopathological images of colon cancer and obtained impressive results. In [35], the authors studied hepatic tumors using MRI images of the liver, using TDA as a feature extraction method from 2D MRI slices. In [12], the authors examined brain cancer by utilizing a sequence of binary images obtained via tumor segmentation, and capturing topological patterns of brain tumors through the Euler Characteristics Curve in different radial directions.

Although these methods use different variations of sub-level filtrations, they all utilize 2D images and thus differ substantially from our approach. 2D images are limited to 0 and 1-dimensional topological features (components and loops), relying solely on these lower-dimensional features for their study. In contrast, our approach employs 3D MRI images (volumes), with key features derived from 2-dimensional topological features that represent cavities in $3D$ cubical complexes.

## 3. Background

### 3.1. Persistent Homology

Persistent homology (PH) studies how the topology of a space changes as we vary a parameter that controls the level of detail or resolution of the space. It is particularly useful for studying data sets that have a natural ordering or hierarchy. The main difference between homology and PH is that homology provides a static snapshot of the topology of a space, while PH captures the evolution of the topology over different scales or levels of detail. In other words, homology is useful for understanding the global structure of a space, while PH is better suited for detecting and characterizing the local features of a space.

The core concept behind PH is to describe the changes in homology that occur as an object evolves with respect to a parameter. Given a simplicial complex, $\Sigma$, we produce a finite sequence of subcomplexes: $\Sigma_f = \{\Sigma_p | 0 \leq p \leq m\}$ such that $\emptyset = \Sigma_0 \subseteq \Sigma_1 \cdots \subseteq \Sigma_m = \Sigma$ using a filtration technique. Figure 1 illustrates the evolution of the homology groups over the entire sequence of simplicial complexes.

### 3.2. Cubical Complex Filtration

In this paper, we present a method for applying PH to 3D MRI images, which involves inducing a nested sequence of 3D cubical complexes and tracking their topological features. We represent a 3D MRI image $\mathcal{X}$ as a 3D cubical complex with resolution $r \times s \times t$. To perform PH, we first create a nested sequence of 3D binary images by considering grayscale values $\gamma_{ijk} \in [0, 255]$ of each voxel $\Delta_{ijk} \subset \mathcal{X}$. For a sequence of grayscale values $(0 \leq t_1 < t_2 < \cdots < t_N \leq 255)$, we obtain 3D binary images $\mathcal{X}_1 \subset \mathcal{X}_2 \subset \cdots \subset \mathcal{X}_N$, where $\mathcal{X}_n$ is the set of voxels $\Delta_{ijk} \subset \mathcal{X}$ with $\gamma_{ijk} \leq t_n$. In other words, we start with an empty 3D image of size $r \times s \times t$, and activate voxels by coloring them black as their grayscale values exceed the threshold, $t_n$, at each step. This process is illustrated in Figure 2. The PH algorithm then tracks the topological features, such as connected components, holes/loops, and cavities, across this sequence of binary images [10, 23]. Overall, this method allows us to gain insights into the complex topological structures of 3D MRI images and can be used for various applications, such as medical image classification and segmentation.

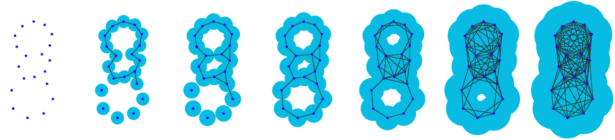

Figure 1. The filtration procedure begins with a set of points and proceeds by incrementally increasing the radius of the balls centered on each point. This procedure constructs a sequence of simplicial complexes that capture the topological features of the space in increasing levels of detail or resolution. Image is from [8].

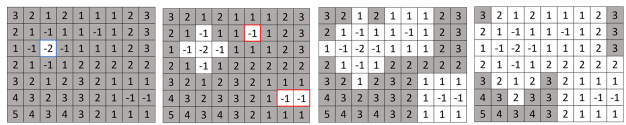

Figure 2. At $t_n \leq -2$, a connected component is born, and at $t_n \leq -1$, two other connected components are born. When $t_n \leq 1$, the upper component with birth-time $-1$ is killed and merged into the one with birth-time $-2$. Finally, when $t_n \leq 2$, the right component with birth-time $-1$ is killed. As a result, the 0-th barcodes of the dataset can be computed and represented as $[-2, \inf)$, $[-1, 1)$, and $[-1, 2)$. This barcode representation provides a concise and informative summary of the topological structure of the dataset. Image is from [23].

PH tracks the evolution of topological features across a sequence of cubical complexes $\mathcal{X}_n$ and records it as a persistence diagram (PD). The PD provides information on the birth and death times of topological features, such as connected components, loops, and voids, which appear and disappear over the sequence. Specifically, if a topological feature $\sigma$ first appears in $\mathcal{X}_m$ and disappears in $\mathcal{X}_n$, we refer to $b_\sigma = t_m$ as the birth time and $d_\sigma = t_n$ as the death time of the feature $\sigma$. The PD for dimension $k$ is then the collection of all such 2-tuples $PD_k(\mathcal{X}) = (b_\sigma, d_\sigma)$, which represents the persistence of the $k$-dimensional topological features across the sequence. This approach enables us to identify the significant topological features in the data and analyze their persistence over different scales or levels of resolution.

## 4. Measuring Pairwise Distances Between Persistence Diagrams

### 4.1. Sinkhorn-Knopp Algorithm

The optimal transport problem seeks to find the most efficient way to transport a set of masses from one distribution to another, where the cost of transporting a mass depends on the distance between its source and destination points. The Wasserstein distance is not explicitly used in the equations of Sinkhorn's algorithm, but it arises naturally from the optimization problem being solved.

In the context of persistence diagrams, the masses correspond to the persistence pairs, and the cost of transporting a pair depends on the distance between its birth and death values. The cost matrix $C$ used in Sinkhorn's algorithm encodes these distances, and the optimal transport plan $P$ determines how to match the pairs in the two diagrams in order to minimize the total cost. Thus, the Wasserstein distance arises implicitly from the optimal transport problem being solved by Sinkhorn-Knopp algorithm. The final output of the algorithm (the optimal transport plan $P$) can be used to compute the Wasserstein distance between the persistence diagrams.

**Definition 4.1.** Doubly stochastic matrix, denoted by $D \in \mathcal{R}^{N \times N}$, is a square matrix with non-negative entries that is both row-normalized and column-normalized:

$$\sum_{i=1}^{N} D_{ij} = 1, \quad \forall j \in [1, N], \quad \sum_{j=1}^{N} D_{ij} = 1, \quad \forall i \in [1, N]$$

**Definition 4.2.** The Sinkhorn-Knopp Algorithm states that any square matrix comprising strictly positive elements can be transformed into $\wedge_1 D \wedge_2$, where $\wedge_1$ and $\wedge_2$ are diagonal matrices with strictly positive diagonal entries.

Sinkhorn-Knopp Algorithm approximates the doubly stochastic matrix with linear convergence. The algorithm operates by iteratively normalizing the matrix along its rows and columns:

```
def sinkhorn(D, N, L):
    # Input: positive matrix D[N x N]
    # L: max. number of iterations
    for i in range(L):
        D = D / np.matmul(D, np.ones(N, 1))
        D = D / np.matmul(np.ones(1, N), D)
        # Test for convergence and early stop.
        if converge:
            break
    return D
```

### 4.2. Optimal Transport Plan

---
**Algorithm 1** Sinkhorn's algorithm for optimal transport
---
**Require:** Persistence diagrams $D_1$ and $D_2$, regularization parameter $\mu$, and $\epsilon$ to avoid the zero division.
**Ensure:** Optimal transport plan $P$.
1: Define the cost matrix $C \in \mathcal{R}^{N \times M}$, where $N$ and $M$ are the number of points in $D_1$ and $D_2$, respectively, and $C_{i,j}$ represents the cost of matching the $i$th point of $D_1$ with the $j$th point of $D_2$.
2: Initialize row normalization vector $u \in \mathcal{R}^{1 \times N}$ and column normalization vector $v \in \mathcal{R}^{1 \times M}$ to be uniform probability distributions over the rows and columns of the cost matrix, respectively:

$$u \leftarrow \frac{\text{ones}(N)}{N} \quad \text{and} \quad v \leftarrow \frac{\text{ones}(M)}{M}$$

3: Define the regularization parameter $\mu$, which controls the trade-off between accuracy and stability:

$$\mu \leftarrow 0.01, \quad \epsilon \leftarrow 10^{-99}, \quad \max_{\text{iter}} \leftarrow 1000$$

4: Compute the exponentiated cost matrix to transform the original linear programming problem into a more computationally tractable form, introducing entropic regularization and enabling an efficient iterative solution:

$$K \leftarrow \exp^{-C/\mu}$$

5: Update row and column normalization vectors:
6: **for** $k = 1$ to $\max_{\text{iter}}$ **do**
7:     **while** not converged **do**

$$u \leftarrow \frac{1}{K \cdot v + \epsilon} \quad \text{and} \quad v \leftarrow \frac{1}{K^\top \cdot u + \epsilon}$$

8:     **end while**
9: **end for**
10: Compute the optimal transport plan $P$, and return the Wasserstein distance between $D_1$ and $D_2$:

$$P \leftarrow \text{diag}(u) \cdot K \cdot \text{diag}(v)$$

$$\text{distance} \leftarrow \left( \sum_{i=1}^{N} \sum_{j=1}^{M} P_{i,j} C_{i,j} \right)$$

11: **return** distance
---

The resulting optimal transport plan $P$ represents the most efficient way to match the points of $D_1$ to the points of $D_2$. This algorithm is used in a variety of applications, including shape matching and image registration. Next, we present a Python code implementation for computing the optimal transport plan between two persistence diagrams utilizing the Sinkhorn-Knopp algorithm and produce the Wasserstein distance with an example.

```python
import numpy as np

def persistence_distance(p1, p2, p=2, tol=1e-6):
    """
    Compute the persistence distance between two
    persistence diagrams p1 and p2.
    Parameters:
    -----------
    p1 (np.array): of shape (n1, 2)
        The first persistence diagram.
    p2 (np.array): of shape (n2, 2)
        The second persistence diagram.
    p (float): optional (default=2)
        The exponent used to compute the
        cost matrix.
    tol (float): The convergence threshold for
        the algorithm. Default is 1e-6.
    Returns:
    --------
    distance (float):
        Wasserstein distance between p1 and p2.
    """
    # compute cost matrix
    n1, n2 = p1.shape[0], p2.shape[0]
    C = np.zeros((n1, n2))
    for i in range(n1):
        for j in range(n2):
            C[i,j] = np.abs(p1[i,0] - p2[j,0])\
            **p + np.abs(p1[i,1] - p2[j,1])**p

    # set algorithm parameters
    reg = 0.01
    epsilon = 1e-99
    max_iter = 1000

    # compute the exponentiated cost matrix
    K = np.exp(-C / reg)

    # row and column normalization vectors
    u = np.ones(n1) / n1
    v = np.ones(n2) / n2

    # run Sinkhorn-Knopp algorithm
    for k in range(max_iter):
        u_prev, v_prev = u.copy(), v.copy()

        # Update u and v
        u = 1 / (np.dot(K, v) + epsilon)
        v = 1 / (np.dot(K.T, u) + epsilon)

        # Check for convergence
        if np.allclose(u, u_prev, rtol=0,
                    atol=tol) and\
          np.allclose(v, v_prev, rtol=0,
                    atol=tol):
                break

    # compute the optimal transport plan
    P = np.diag(u) @ K @ np.diag(v)

    # return the persistence distance
    return np.sum(np.multiply(P, C))
```

We can provide an illustrative example to demonstrate the efficacy of our approach. Let's define 2 persistence diagrams as a NumPy array, where the first column corresponds to the birth time and the second column denotes the death time of a topological feature,

```
p1 = np.array([[1, 2], [3, 4], [5, 6]])
p2 = np.array([[2, 3], [4, 5], [6, 7]])
```

We obtain the resultant cost matrix, $C$ as follows,

```
[[ 2. 18. 50.]
 [ 2.  2. 18.]
 [18.  2.  2.]]
```

The optimal transport plan matrix, $P \in \mathcal{R}^{N \times M}$ represents the optimal way to match the points in the two input persistence diagrams, $p_1$ and $p_2$, in order to minimize the cost of the matching. Each element $P_{i,j}$ of the matrix represents the amount of mass that should be transported from point $i$ in $p_1$ to point $j$ in $p_2$, in order to minimize the overall cost of the matching. $P$ is subject to row-wise and column-wise normalization. These constraints maintain the conservation of mass throughout the matching process, as they ensure that the total mass remains constant in each row and column.

```
[[0.999 0.000 0.000]
 [0.001 0.999 0.000]
 [0.000 0.001 0.999]]
```

Ultimately, the computed Wasserstein distance yields a value of 6.0.

### 4.3. Topology-Aware Focal Loss

Our approach attains per-pixel precision as well as topological correctness through the utilization of topology-aware focal loss (TAFL), $L_{tafl}(f, g)$ by training a 3D U-Net for segmentation. $f$ denotes the predicted segmentation mask, while $g$ represents the ground truth. The training loss consists of a weighted combination of per-pixel focal loss, $L_{focal}$ and the aforementioned topological loss, $L_{topo}$ that is the elementwise sum over the Hadamard product of optimal transport plan matrix, $P$, and cost matrix, $C$.

$$L_{tafl}(f,g) = L_{focal}(f,g) + \lambda L_{topo}(f,g)$$
$$L_{focal}(f,g) = -\alpha_t(1-p_t)^\gamma \log(p_t)$$
$$L_{topo}(f,g) = \left(\sum_{i=1}^{N}\sum_{j=1}^{M} P_{i,j} C_{i,j}\right)$$

where $\lambda$ is the regularization parameter that modulates the weight of topological loss. In our experiments, we have found that setting $\lambda$ equal to 0.001 yield optimal stability in loss convergence. Additionally, $\alpha$ is the weighting factor for focal loss and set to $1$ and $\gamma$ is the scaling factor that addresses the issue of class imbalance in image segmentation tasks, whereby well-classified samples, $p_t > 0.5$, are given a lower loss while misclassified samples are emphasized more. We have set the value of $\gamma$ to 2 in our experiments. The probability estimate for each class is denoted by $p_t$.

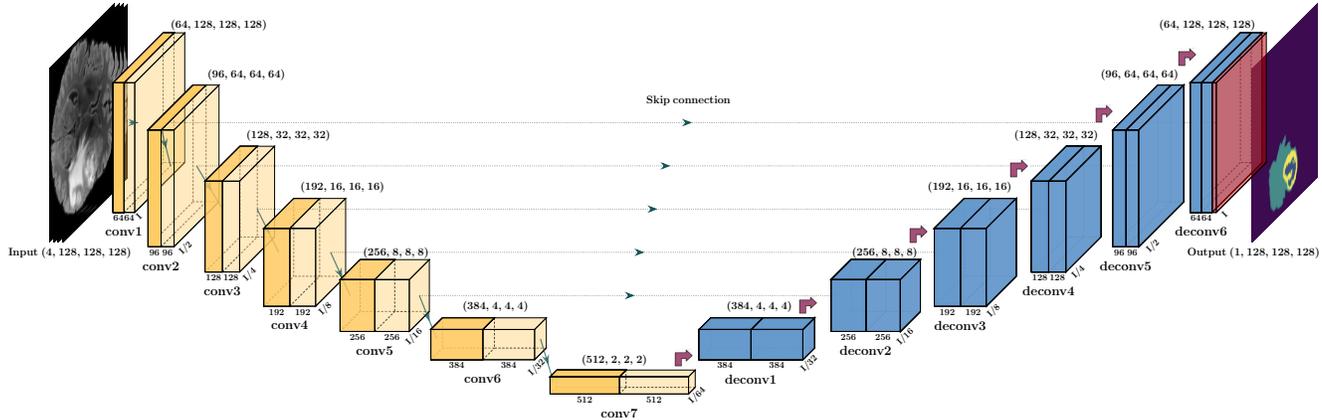

Figure 3. 3D U-Net Architecture. The U-Net design employs an encoder module that applies spatial dimension reduction to the input, which is subsequently upsampled by the decoder module to restore the input to its original shape.

## 5. Experiments

### 5.1. Datasets

**BraTS 2019** dataset[1] consists of 335 3D brain MRI scans, segmented manually by one to four experienced neuroradiologists ensuring consistency in the tumor annotation protocol [3, 4, 27, 29]. 259 of them belong to patients diagnosed with HGG and 76 of them to those diagnosed with LGG. Segmentation annotations comprise four classes: the necrotic and non-enhancing tumor core (NCR/NET–label 1), the peritumoral edema (ED–label 2), background (voxels that are not part of the tumor–label 3) and GD-enhancing tumor (ET–label 4). Each MRI scan image has 4 modalities: native (T1) and post-contrast T1-weighted (T1Gd), T2-weighted (T2), and d) T2 Fluid Attenuated Inversion Recovery (T2-FLAIR) and each modality has dimensions of (240, 240, 155) voxels. The provided data is skull-stripped and interpolated to the same resolution of 1 mm$^3$.

### 5.2. Preprocessing

Data augmentation transforms were applied to the batches of data sampled from training set to increase the robustness of the U-Net segmentation model and learning a higher quality latent-space representation. In order to mitigate the overfitting problem while training U-Net for tumor segmentation, we synthetically increase the size of the training set by a composition of transforms, which includes: random cropping, rotating, injecting Gaussian noise, adding Gaussian blur, changing brightness and contrast. Specifically, we randomly crop patches of size $(5, 128, 128, 128)$ per MRI scan id, where 5 denotes 4 MRI modalities and their corresponding segmentation mask. Then we apply random flipping along the x, y and z axes with a probability of 0.5. This is followed by injecting Gaussian noise (with mean, $\mu = 0$ and standard deviation, $\sigma \sim \mathcal{U}[0, 0.33]$) with a probability of 0.2 and adding Gaussian blur (where $\sigma$ of the Gaussian kernel is sampled from a uniform distribution: $\sigma \sim \mathcal{U}[0.5, 1.5]$) with a probability of 0.2. We jitter brightness with a brightness factor uniformly sampled from $\mathcal{U}[0.7, 1.4]$ and contrast with a contrast factor uniformly sampled from $\mathcal{U}[0.7, 1.4]$ at probability of 0.2.

Let $f(x)$ denote a function approximator, which is a CNN, and $g(x)$ an image transform, specifically translation that shifts a voxel in the 3D volume such that $I(x, y, z) = I(x-a, y-b, z-c)$, where $I(x, y, z)$ denotes the voxels in the 3D image and $a, b, c$ are scalars. $f(x)$ is said to be equivariant with respect to $g(x)$, if $f(g(x)) = g(f(x))$. CNNs are inherently equivariant with respect to translation thanks to the concept of weight sharing, however they are not naturally equivariant to other transformations such as changes in scale, rotation or perturbation with noise. Utilizing translational equivariance by applying transformations such as flipping, zooming or translating in conjunction with noise perturbation and color jittering on each batch of data during training phase comes handy to increase the performance of CNNs and make them more robust against adversaries. However, data augmentation transforms are not expected to achieve the desired outcome while producing persistent homology features, since persistent homology already generates topological features that persist across several scales. An important property of topological features is that they are homotopy invariant as they do not change, when the underlying space is transformed by stretching, bending or other deformations [34]. Hence, we exempt the training data from augmentation transforms while operating cubical

---
[1] https://www.med.upenn.edu/cbica/brats2019/data.html

persistent homology to ease the computational complexity.

## 5.3. Macro Design

Figure 3 shows that the U-Net architecture is composed of 2 parts, i.e., encoder and decoder. Encoder has 7 convolutional blocks, where each block consists of 2 repeating convolutional layers. The first layer applies Conv3d kernels with size $(3 \times 3 \times 3)$ and stride $(2, 2, 2)$, followed by InstanceNorm3d and GeLU activation, whereas the second layer applies kernels of size $(3 \times 3 \times 3)$ and stride $(1, 1, 1)$, also followed by InstanceNorm3d and GeLU. We adopt a "patchify" strategy, and eliminate the pooling layer by applying convolutional kernels with stride $(2, 2, 2)$, which effectively yields a $(2, 2, 2)$ downsampling while preserving shift-equivariance. Furthermore, we replace convex and monotonic ReLU activation with a non-convex and non-monotonic activation function GeLU, which is differentiable in its entire domain and allows non-zero gradients for negative inputs unlike ReLU.

After generating a latent vector, ConvTranspose3d operators of kernel size $(2 \times 2 \times 2)$ and stride $(2, 2, 2)$, represented with purple arrows in Figure 3, increases the spatial dimensions of the output from previous layer by a factor of 2 until the spatial dimensions of the decoder's output is equivalent to the original input image. An upsampled feature map is concatenated with the encoder feature map of the equivalent spatial level (except of those between conv6 and deconv1 blocks). Symmetrical long skip connections between encoder and decoder blocks, represented with green, dotted lines, enhances the reconstruction of fine-grained details. As far as we know there is no previous research showing the theoretical justification of why long skip connections improve the reconstruction performance, but we conjecture that these lengthy skip connections utilized to transmit the feature maps from early encoder layers to late decoder layers allow to recover the spatial information that is lost during downsampling. The output feature map is passed to a decoder block which consists of 2 repeating, identical, Conv3d layers with kernel size $(3 \times 3 \times 3)$ and stride $(1, 1, 1)$. Similary, each Conv3d layer is followed by InstanceNorm3d and GeLU.

To enhance the robustness of our predictions, a well-established approach involves the implementation of test time augmentations. Specifically, during the inference process, we generate multiple variants of the input volume by applying one of eight potential flips along the x, y, and z axes. For each of these versions, we perform inference and subsequently transform the resulting predictions back to the original orientation of the input volume by utilizing the same flips as those employed during input volume processing. Ultimately, we aggregate the probabilities derived from all predictions by computing their average.

Table 1. The average Dice scores for the ET, TC and WT classes, as well as for all classes combined.

| Loss Objective | ET Dice | TC Dice | WT Dice | Dice |
|---|---|---|---|---|
| Cross-Entropy | 0.7255 | 0.7240 | 0.8479 | 0.9570 |
| Focal | 0.7244 | 0.7405 | 0.8601 | 0.9582 |
| **TAFL** | **0.7383** | **0.8005** | **0.8793** | **0.9602** |

## 5.4. Training Schedule

Our approach involves training models from scratch, with no use of fine-tuning. We do not utilize any pre-existing models as a starting point and initialize the weights of our model randomly. This allows us to fully explore the model's parameter space, without the bias or constraints imposed by a pre-existing model. We utilized the Adam optimizer with a learning rate of 0.0005 and a weight decay set to 0.0001 for 200 epochs. To ensure effective training, we implemented a cosine annealing scheduler to dynamically adjust the learning rate throughout the training process. We saved the checkpoint with the highest mean Dice score achieved on the validation set during training.

## 5.5. Empirical Evaluations

ET Dice score quantifies the extent of intersection between the ground truth and the segmented enhancing tumor region. WT amalgamates the edema, enhancing tumor and necrotic tumor core classes to evaluate the degree of overlap. Finally, TC employs the segmented necrotic tumor to evaluate the degree of similarity between the ground truth and the predicted segmentations. The experimental results presented in Table 1 show that training with topology-aware focal loss outperforms training with cross-entropy and focal loss. To ensure fair comparison, we follow a standardized evalution process by using the same training and validation dataset, same U-Net architecture and epoch size. Please note that in the context of brain tumor segmentation, a good Dice score would depend on various factors such as the complexity of the tumor structure, the size and shape of the tumors, and the quality of the input data. Generally, a Dice score of $0.8$ or higher is considered to be a good performance for brain tumor segmentation. See Figure 4 for the segmentation results recorded on the validation set.

## 6. Conclusion

We have introduced a novel loss function to address topological errors that frequently arise in segmentation algorithms. We evaluated TAFL on the BraTS challenge validation dataset, and the results showed that it enhances the quality of segmentation performance. Our future work involves investigating the generalization of TAFL to other segmentation tasks and extending its applicability to a wider range of medical imaging modalities.

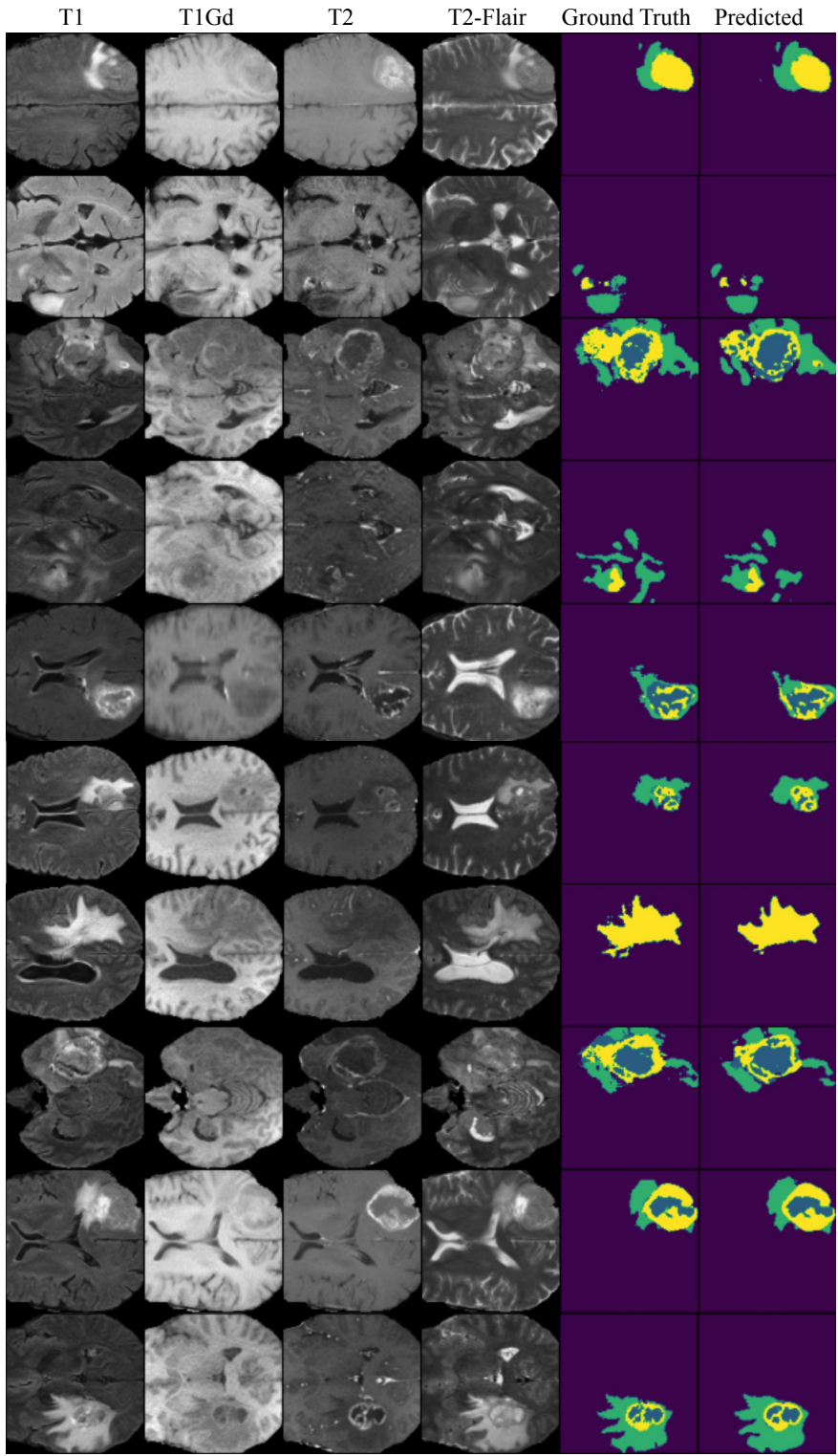

Figure 4. The visualization of segmentation outcomes involves the display of a 3D brain MRI scan image slice featuring modalities T1, T1Gd, T2, and T2-Flair, as well as both ground truth and predicted segmentation masks. To aid in interpretation, color coding is utilized, with purple representing the background or healthy cells, green representing peritumoral edema, yellow indicating an enhancing tumor, and blue representing the necrotic and non-enhancing tumor core.

# References

[1] Aaron Adcock, Daniel Rubin, and Gunnar Carlsson. Classification of hepatic lesions using the matching metric. *Computer vision and image understanding*, 121:36–42, 2014. 2

[2] Vijay Badrinarayanan, Alex Kendall, and Roberto Cipolla. Segnet: A deep convolutional encoder-decoder architecture for image segmentation. *IEEE transactions on pattern analysis and machine intelligence*, 39(12):2481–2495, 2017. 1

[3] S Bakas, H Akbari, A Sotiras, M Bilello, M Rozycki, J Kirby, J Freymann, K Farahani, and C Davatzikos. Segmentation labels for the pre-operative scans of the tcga-gbm collection. the cancer imaging archive, 2017. 6

[4] Spyridon Bakas, Mauricio Reyes, Andras Jakab, Stefan Bauer, Markus Rempfler, Alessandro Crimi, Russell Takeshi Shinohara, Christoph Berger, Sung Min Ha, Martin Rozycki, et al. Identifying the best machine learning algorithms for brain tumor segmentation, progression assessment, and overall survival prediction in the brats challenge. *arXiv preprint arXiv:1811.02629*, 2018. 6

[5] Paul Bendich, James S Marron, Ezra Miller, Alex Pieloch, and Sean Skwerer. Persistent homology analysis of brain artery trees. *The annals of applied statistics*, 10(1):198, 2016. 2

[6] Stan Benjamens, Pranavsingh Dhunnoo, and Bertalan Meskó. The state of artificial intelligence-based fda-approved medical devices and algorithms: an online database. *NPJ digital medicine*, 3(1):1–8, 2020. 2

[7] Eric Berry, Yen-Chi Chen, Jessi Cisewski-Kehe, and Brittany Terese Fasy. Functional summaries of persistence diagrams. *Journal of Applied and Computational Topology*, 4(2):211–262, 2020. 2

[8] Peter Bubenik et al. Statistical topological data analysis using persistence landscapes. *J. Mach. Learn. Res.*, 16(1):77–102, 2015. 3

[9] Pablo G Cámara, Arnold J Levine, and Raul Rabadan. Inference of ancestral recombination graphs through topological data analysis. *PLoS computational biology*, 12(8):e1005071, 2016. 2

[10] Seungho Choe and Sheela Ramanna. Cubical homology-based machine learning: An application in image classification. *Axioms*, 11(3):112, 2022. 3

[11] Elizabeth B Claus, Kyle M Walsh, John K Wiencke, Annette M Molinaro, Joseph L Wiemels, Joellen M Schildkraut, Melissa L Bondy, Mitchel Berger, Robert Jenkins, and Margaret Wrensch. Survival and low-grade glioma: the emergence of genetic information. *Neurosurgical focus*, 38(1):E6, 2015. 2

[12] Lorin Crawford, Anthea Monod, Andrew X Chen, Sayan Mukherjee, and Raúl Rabadán. Predicting clinical outcomes in glioblastoma: an application of topological and functional data analysis. *Journal of the American Statistical Association*, 115(531):1139–1150, 2020. 2, 3

[13] Olga Dunaeva, Herbert Edelsbrunner, Anton Lukyanov, Michael Machin, Daria Malkova, Roman Kuvaev, and Sergey Kashin. The classification of endoscopy images with persistent homology. *Pattern Recognition Letters*, 83:13–22, 2016. 2

[14] Michał Futrega, Alexandre Milesi, Michał Marcinkiewicz, and Pablo Ribalta. Optimized u-net for brain tumor segmentation. In *International MICCAI Brainlesion Workshop*, pages 15–29. Springer, 2022. 2

[15] Kathryn Garside, Robin Henderson, Irina Makarenko, and Cristina Masoller. Topological data analysis of high resolution diabetic retinopathy images. *PloS one*, 14(5):e0217413, 2019. 2

[16] Noah Giansiracusa, Robert Giansiracusa, and Chul Moon. Persistent homology machine learning for fingerprint classification. In *2019 18th IEEE International Conference On Machine Learning And Applications (ICMLA)*, pages 1219–1226. IEEE, 2019. 2

[17] Barbara Giunti. Tda applications library, 2022. https://www.zotero.org/groups/2425412/tda-applications/library. 2

[18] Ali Hatamizadeh, Yucheng Tang, Vishwesh Nath, Dong Yang, Andriy Myronenko, Bennett Landman, Holger R Roth, and Daguang Xu. Unetr: Transformers for 3d medical image segmentation. In *Proceedings of the IEEE/CVF Winter Conference on Applications of Computer Vision*, pages 574–584, 2022. 2

[19] Kaiming He, Georgia Gkioxari, Piotr Dollár, and Ross Girshick. Mask r-cnn. In *Proceedings of the IEEE international conference on computer vision*, pages 2961–2969, 2017. 1

[20] Kaiming He, Xiangyu Zhang, Shaoqing Ren, and Jian Sun. Deep residual learning for image recognition. In *Proceedings of the IEEE conference on computer vision and pattern recognition*, pages 770–778, 2016. 2

[21] Christoph Hofer, Roland Kwitt, Marc Niethammer, and Andreas Uhl. Deep learning with topological signatures. *arXiv preprint arXiv:1707.04041*, 2017. 1

[22] Xiaoling Hu, Fuxin Li, Dimitris Samaras, and Chao Chen. Topology-preserving deep image segmentation. *Advances in neural information processing systems*, 32, 2019. 1

[23] Shizuo Kaji, Takeki Sudo, and Kazushi Ahara. Cubical ripser: Software for computing persistent homology of image and volume data. *arXiv preprint arXiv:2005.12692*, 2020. 3

[24] Lida Kanari, Paweł Dłotko, Martina Scolamiero, Ran Levi, Julian Shillcock, Kathryn Hess, and Henry Markram. A topological representation of branching neuronal morphologies. *Neuroinformatics*, 16(1):3–13, 2018. 2

[25] Anita Khanna, Narendra D Londhe, S Gupta, and Ashish Semwal. A deep residual u-net convolutional neural network for automated lung segmentation in computed tomography images. *Biocybernetics and Biomedical Engineering*, 40(3):1314–1327, 2020. 2

[26] Michel Lacroix, Dima Abi-Said, Daryl R Fourney, Ziya L Gokaslan, Weiming Shi, Franco DeMonte, Frederick F Lang, Ian E McCutcheon, Samuel J Hassenbusch, Eric Holland, et al. A multivariate analysis of 416 patients with glioblastoma multiforme: prognosis, extent of resection, and survival. *Journal of neurosurgery*, 95(2):190–198, 2001. 2

[27] Christopher T Lloyd, Alessandro Sorichetta, and Andrew J Tatem. High resolution global gridded data for use in population studies. *Scientific data*, 4(1):1–17, 2017. 6

[28] Ephy R Love, Benjamin Filippenko, Vasileios Maroulas, and Gunnar Carlsson. Topological deep learning. *arXiv preprint arXiv:2101.05778*, 2021. 1

[29] Jakab Menze, Himmelreich Masuch, Petrich Bachert, et al. Menze bh. *Kelm BM, Masuch R., Himmelreich U., Bachert P., Petrich W., et al., A comparison of random forest and its gini importance with standard chemometric methods for the feature selection and classification of spectral data, BMC Bioinformatics*, 10(1), 2009. 6

[30] Agata Mosinska, Pablo Marquez-Neila, Mateusz Koziński, and Pascal Fua. Beyond the pixel-wise loss for topology-aware delineation. In *Proceedings of the IEEE conference on computer vision and pattern recognition*, pages 3136–3145, 2018. 1

[31] Andriy Myronenko. 3d mri brain tumor segmentation using autoencoder regularization. In *International MICCAI Brainlesion Workshop*, pages 311–320. Springer, 2018. 2

[32] Monica Nicolau, Arnold J Levine, and Gunnar Carlsson. Topology based data analysis identifies a subgroup of breast cancers with a unique mutational profile and excellent survival. *Proceedings of the National Academy of Sciences*, 108(17):7265–7270, 2011. 1

[33] Ozan Oktay, Jo Schlemper, Loic Le Folgoc, Matthew Lee, Mattias Heinrich, Kazunari Misawa, Kensaku Mori, Steven McDonagh, Nils Y Hammerla, Bernhard Kainz, et al. Attention u-net: Learning where to look for the pancreas. *arXiv preprint arXiv:1804.03999*, 2018. 1, 2

[34] Nina Otter, Mason A Porter, Ulrike Tillmann, Peter Grindrod, and Heather A Harrington. A roadmap for the computation of persistent homology. *EPJ Data Science*, 6:1–38, 2017. 6

[35] Asuka Oyama, Yasuaki Hiraoka, Ippei Obayashi, Yusuke Saikawa, Shigeru Furui, Kenshiro Shiraishi, Shinobu Kumagai, Tatsuya Hayashi, and Jun'ichi Kotoku. Hepatic tumor classification using texture and topology analysis of non-contrast-enhanced three-dimensional t1-weighted mr images with a radiomics approach. *Scientific reports*, 9(1):1–10, 2019. 2, 3

[36] Talha Qaiser, Yee-Wah Tsang, Daiki Taniyama, Naoya Sakamoto, Kazuaki Nakane, David Epstein, and Nasir Rajpoot. Fast and accurate tumor segmentation of histology images using persistent homology and deep convolutional features. *Medical image analysis*, 55:1–14, 2019. 2, 3

[37] Bastian Rieck, Tristan Yates, Christian Bock, Karsten Borgwardt, Guy Wolf, Nicholas Turk-Browne, and Smita Krishnaswamy. Uncovering the topology of time-varying fmri data using cubical persistence. *Advances in neural information processing systems*, 33:6900–6912, 2020. 2

[38] Olaf Ronneberger, Philipp Fischer, and Thomas Brox. U-net: Convolutional networks for biomedical image segmentation. In *International Conference on Medical image computing and computer-assisted intervention*, pages 234–241. Springer, 2015. 1, 2

[39] Madeleine M Shaver, Paul A Kohanteb, Catherine Chiou, Michelle D Bardis, Chanon Chantaduly, Daniela Bota, Christopher G Filippi, Brent Weinberg, Jack Grinband, Daniel S Chow, et al. Optimizing neuro-oncology imaging: a review of deep learning approaches for glioma imaging. *Cancers*, 11(6):829, 2019. 1

[40] Nahian Siddique, Sidike Paheding, Colin P Elkin, and Vijay Devabhaktuni. U-net and its variants for medical image segmentation: A review of theory and applications. *Ieee Access*, 9:82031–82057, 2021. 2

[41] Yara Skaf and Reinhard Laubenbacher. Topological data analysis in biomedicine: A review. *Journal of Biomedical Informatics*, page 104082, 2022. 3

[42] Primoz Skraba, Maks Ovsjanikov, Frederic Chazal, and Leonidas Guibas. Persistence-based segmentation of deformable shapes. In *2010 IEEE Computer Society Conference on Computer Vision and Pattern Recognition-Workshops*, pages 45–52. IEEE, 2010. 2

[43] Cancer Genome Atlas Research Network Tissue source sites: Duke University Medical School McLendon Roger 1 Friedman Allan 2 Bigner Darrell 1, Emory University Van Meir Erwin G. 3 4 5 Brat Daniel J. 5 6 M. Mastrogianakis Gena 3 Olson Jeffrey J. 3 4 5, Henry Ford Hospital Mikkelsen Tom 7 Lehman Norman 8, MD Anderson Cancer Center Aldape Ken 9 Alfred Yung WK 10 Bogler Oliver 11, University of California San Francisco VandenBerg Scott 12 Berger Mitchel 13 Prados Michael 13, et al. Comprehensive genomic characterization defines human glioblastoma genes and core pathways. *Nature*, 455(7216):1061–1068, 2008. 2

[44] Bernadette J Stolz, Tegan Emerson, Satu Nahkuri, Mason A Porter, and Heather A Harrington. Topological data analysis of task-based fmri data from experiments on schizophrenia. *Journal of Physics: Complexity*, 2(3):035006, 2021. 2

[45] Run Su, Deyun Zhang, Jinhuai Liu, and Chuandong Cheng. Msu-net: Multi-scale u-net for 2d medical image segmentation. *Frontiers in Genetics*, 12:639930, 2021. 1

[46] Michael Weller, Martin van den Bent, Matthias Preusser, Emilie Le Rhun, Jörg C Tonn, Giuseppe Minniti, Martin Bendszus, Carmen Balana, Olivier Chinot, Linda Dirven, et al. Eano guidelines on the diagnosis and treatment of diffuse gliomas of adulthood. *Nature reviews Clinical oncology*, 18(3):170–186, 2021. 2